# Zero-Knowledge Games

Ian Malloy

Abstract: In this paper we model a game such that all strategies are non-revealing, with imperfect recall and incomplete information. We also introduce a modified sliding-block code as a linear transformation which generates common knowledge of how informed a player is. Ultimately, we see that between two players in a zero-knowledge game where both players are informed, the utility of trust is established in the mixed strategy Nash equilibrium. Of note is the "will to verify" where a player may be non-cooperative, uninformed, or unfaithful. A zero-knowledge game is one of trust and soundness, placing utility in being informed. For any player who may be uninformed, such players reveal they are uninformed. Additionally, we introduce a strategy polytope embedded within the strategy simplex as a certain case of F-N-Nash equilibrium.

## Introduction

Games are defined as the study of information and linguistic mediums in strategic communication (Arfi 2006). Game theory is also the potential for, and possibility of, communication and coalitions (Nash 1950). If such games are zero-knowledge then the utility becomes trust and is gained by being informed. A zero-knowledge game isolates propositions and their proofs as distinct areas to challenge. Zero-knowledge proofs are receiving heightened attention as protocols capable of securing identity, transactions, and offer a unique perspective in game theory. There are several works applying zero-knowledge proofs to blockchain technologies (Zhou, et al. 2024). We define a zero-knowledge game similar to the protocol, with the principles of commitment (Damgard 1998) and non-revealing strategies (Epsen 2006), (Damgard 1998).

The strategies are elements of the set $\prod_{i \in N} S_i$ with probability measures over a finite set $W$, denoted Δw for mixed strategy Nash equilibrium: $\{\Delta w \mid \Delta w \in \Pi_i h(\omega),\ \Delta w \in [0,1]\}$. For a non-empty set $\Omega$, there exists one element of the set wherein the state $\omega \in \Omega$ is true. Treating $\omega$ as the state of a given agent, the extension of this state to a doxastic projection of possible states the agent is in will be symbolized as $h(\omega)$. This is drawn from Jonathan Levin, who describes the state $h(\omega)$ as an agent considering they are in $h(\omega)$ when deciding from the actual state $\omega$ (Levin 2006). The variable $h$ itself is random.

We classify a pair of algorithms $(P, V)$ as Epsen does and follow his definition of a zero-knowledge proof such that $\{\psi \in \Psi, \phi \in \Phi\}$ are strings and $\Sigma^*$ is a language where a protocol implemented in $(P, V)$ results in $V'$ with relative machine $R$ running in polynomial time: then the probability distribution is indistinguishable between $V'$ and $P$ given $(P, V')$ (Epsen 2006). We consider a game between Alice and Bob such that Alice is the verifier and Bob is the prover. For a zero-knowledge game, the principle components are the proposition and its proof. Insofar as the verifier V learns nothing of the string provided by the prover P, the proof is zero-knowledge (Epsen 2006). We assume each game is repeated and each player is ignorant of their payoffs under the same constraints as Epsen (Epsen 2006).

Under Epsen, the repeated game had stage games: $\{G_1, \dots, G_k\}$ and each stage has a relative $m \times n$ matrix of which the entries are known to Alice and Bob (Epsen 2006). After one round, Alice as the verifier challenges Bob the prover and payoffs are calculated. Each challenge and response is $m \times n$ and forms an $n \times n$ matrix with elements of strategies (Epsen 2006). Here we consider all games of the form $\Gamma = \langle N | \prod_{i \in N} S_i \,|\theta\rangle$ with *N*-players, $\prod_{i \in N} S_i$ mixed strategies and $\theta$ payoffs that are indefinite, have finite subgames with repeated moves, imperfect recall, and incomplete information. We further define $\langle \alpha, \beta \rangle$ vectors composed of elements that are confidence levels that a player is informed regarding $\langle \alpha, \beta \rangle$ respectively: $\{\alpha, \beta \mid (\alpha, \beta) \in \mathbb{R},\ \alpha, \beta \in [0,1]\}$. We also regard $\alpha, \beta$ as strategies relative to a challenge and response. Choosing $\alpha$ to challenge leads to a response of either $\alpha$ or $\beta$ that shows how informed they are relative to a proposition $\alpha$ and a proof $\beta$. This necessitates $\alpha, \beta \in \prod_{i \in N} S_i$ while the numerical value assigned to $\alpha, \beta$ is $l_k \in L_k$ as a feasibility level of trust that they are informed relative to $\alpha, \beta$ . While there are *n* players, zero-knowledge proofs require an exchange between at least two parties.

We define every game as common interest, given payoffs are equal between all players (Boors, et al. 2022). Furthermore, each Nash Equilibrium (NE) is also an Altruistic Equilibrium (AE) (Boors, et al. 2022). Not only can any game be decomposed into a zero-sum game, they can also be decomposed into a common interest game (Boors, et al. 2022). In any instance where Alice and Bob wish to engage in an interactive series of negotiations relative to trust in claims and being informed, the game can be decomposed into a zero-knowledge game insofar as non-revealing strategies are possible. Assuming a cost (time, or otherwise) for uninformed players, incurring cost could be the goal.

## Zero-Knowledge Model

Let $\Gamma_1$ be a game: $\Gamma_1 = \langle N | \prod_{i \in N} S_i \, | \theta \rangle$ with *N*-players, $\prod_{i \in N} S_i$ mixed strategies and $\theta$ payoffs.

$$X > Y, \qquad (X, Y) \in \theta$$

| $\Gamma_1$ | $\alpha$ | $\beta$ |
|---|---|---|
| $\alpha$ | X | Y |
| $\beta$ | Y | X |

*Figure 1 - Game One*

The mixed strategy NE is then:

$$\left\{ \Delta w \mid \Delta w = \left( \frac{1}{2}, \frac{1}{2} \right), \ \ \Delta w \in \prod_{i \in N} h(\omega) \right\}$$

Theorem 1: There exists a Markov model of $\Gamma_1$.

Proof:

Let $\mathcal{M} = \begin{bmatrix} \alpha_1 & \alpha_2 \\ \beta_1 & \beta_2 \end{bmatrix}$ such that $\alpha + \beta = 1$ and the following probabilities are true: $P(\alpha|\beta) = P(\beta|\alpha) = 1$.

Assume $\mathcal{M} = \langle f_1(\alpha, \beta), \ -f_1{}'(\alpha, \beta) \rangle$ has linearly independent basis vectors $\langle e_0, e_1 \rangle$ and

$$f_1, f_1{}' \subseteq \prod_{i \in N} S_i$$

Then $\mathcal{M}$ is composed of all permutations of $\langle f_1(\alpha, \beta), \ -f_1{}'(\alpha, \beta) \rangle = \langle e_0, e_0^{-1}, e_1, e_1^{-1} \rangle, \ \{\alpha \perp \beta\}$

Given this$\{L \cap \mathcal{M}\} \subseteq \prod_{i \in N} S_i$ and $L$ forms the strategy simplex and is thus formed of undominated strategies (Nash 1950).

$|\mathcal{M}| = L$ satisfies $\alpha + \beta = 1$ since it is left-stochastic and satisfies $P(\alpha|\beta) = P(\beta|\alpha) = 1$.

The following properties of partitions given by Levin (2006) must be satisfied:

$$\omega \in h(\omega), \qquad \forall \omega \in \Omega$$

$$[\omega' \in h(\omega)] \rightarrow [h(\omega') = h(\omega)]$$

Each state $\omega$ treated by the associative function $h(\omega)$ informs the agent as to the current state with respect to $\omega$ being the actual state. Sampling of the strategy simplex is justified by Nash (Nash 1950).

$M \subseteq \prod_{i \in N} h(\omega)$ 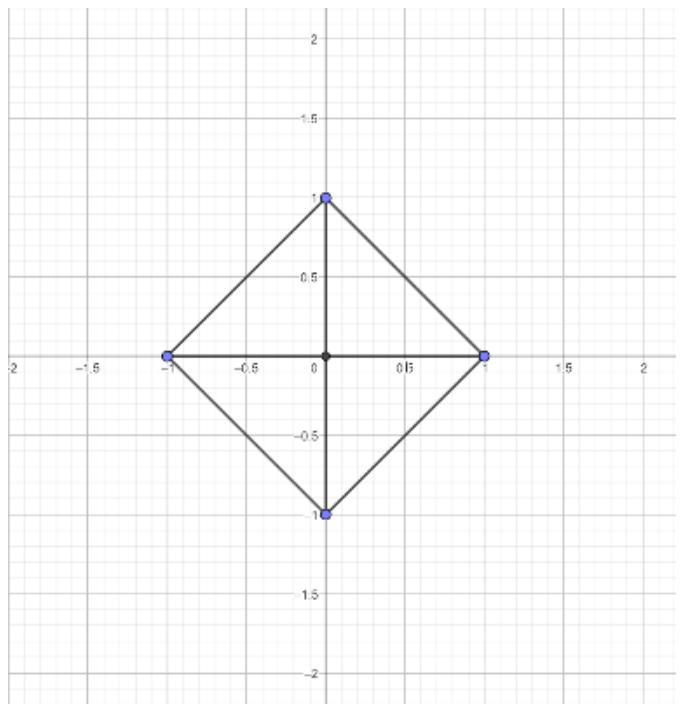 we may then symbolize $L = \begin{bmatrix} 1 & 0 \\ 0 & 1 \end{bmatrix}, \begin{bmatrix} 0 & 1 \\ 1 & 0 \end{bmatrix}$ ■

*Figure 2 - M Strategy Simplex*

Let $\Gamma_1$ be a game $\Gamma_1 = \langle N | \prod_{i \in N} S_i \, | \theta \rangle$ with imperfect recall and incomplete information.

Theorem 2: $\Gamma_1$ has Nash equilibria

Proof: Let Alice's moves be $\alpha, \beta$ and Bob's moves be $\alpha, \beta$. Alice forms an estimate $h_0$ as to whether Bob is informed with respect to $\alpha, \beta$ and Alice then challenges the opposite of her estimate and Bob responds.

$$\langle \alpha, \beta \rangle = \begin{bmatrix} 1 \\ 0 \end{bmatrix}, \begin{bmatrix} 0 \\ 1 \end{bmatrix} \text{ ordered}$$

$\langle h_0 \rangle \rightarrow$ challenge and response $\langle h_0, h_1 \rangle$ symbolized as the operator $\otimes$:

$$h_1 = \begin{cases} \alpha \otimes \alpha = \alpha \\ \beta \otimes \beta = \beta \\ \alpha \otimes \beta = \begin{bmatrix} 0 \\ 0 \end{bmatrix} \\ \beta \otimes \alpha = \begin{bmatrix} 0 \\ 0 \end{bmatrix} \end{cases}$$

$$L = \langle h_{k-1}, h_k \rangle, \qquad \{k | k > 0,\ k \in \mathbb{Z}^+\}$$

Here we introduce the linear transformation $\lambda_k$ as a modified sliding-block code:

$$\lambda_k : L^k = \langle h_{k-1}, \qquad L_{k-1} \cdot h_k \rangle$$

$$L_{k-1} \overset{\lambda}{\rightarrow} L_k$$

$$L_k = \begin{bmatrix} 1 & 0 \\ 0 & 1 \end{bmatrix}, \qquad \begin{bmatrix} 0 & 1 \\ 1 & 0 \end{bmatrix}, \qquad \begin{bmatrix} 0 & 0 \\ 1 & 1 \end{bmatrix}, \qquad \begin{bmatrix} 1 & 1 \\ 0 & 0 \end{bmatrix}$$

We discard those that do not form a simplex in $\mathbb{R}^2$ by sampling a face of the strategy simplex. In this manner, we ensure $L_k$ is undominated when sampling from $L_k$ in one dimension, by being a strategy and forming a simplex in $\mathbb{R}^2$. If $\alpha \otimes \alpha$ or $\beta \otimes \beta$ repeatedly such that it is essentially $\begin{bmatrix} 0 & 0 \\ 1 & 1 \end{bmatrix}, \begin{bmatrix} 1 & 1 \\ 0 & 0 \end{bmatrix}$ then no new information is being provided. Even if $\alpha \otimes \alpha$ or $\beta \otimes \beta$ and we are confident they are informed, never deviating from either the proposition or proof exclusively results in vacuous proofs or repeated instances of propositions of what they claim to prove. Given the exclusive disjunction between the two in cases of $\begin{bmatrix} 0 & 0 \\ 1 & 1 \end{bmatrix}, \begin{bmatrix} 1 & 1 \\ 0 & 0 \end{bmatrix}$ it makes sense to regard this as tautologous.

Removing the tautologies, we then have:

$$L_k = \begin{cases} \begin{bmatrix} 1 & 0 \\ 0 & 1 \end{bmatrix} \\ \begin{bmatrix} 0 & 1 \\ 1 & 0 \end{bmatrix} \end{cases} \subseteq \prod_{i \in N} h(\omega)$$

Furthermore, this $L_k$ restricted to sampling from $\mathcal{M}$ has the associated payoff $\theta = X$ even when payoffs are unknown, and is therefore Nash equilibria given $X > Y$. ∎

Theorem 3:

$$\left|\left|L_k\right|\right|^2 \in \prod_{i \in N} h(\omega)\,, \qquad \left\{\lim_{k\to\infty} \left|\mathcal{M}^k\right| \in \prod_{i \in N} h(\omega) \mid\ (\alpha, \beta) \in \prod_{i \in N} S_i\right\}$$

Proof:

Given Euclidean-norm squared: $\left|\left|L_k\right|\right|^2 \subseteq \left|\left|\mathcal{M}_k\right|\right|^2$ we compute $\lim_{k\to\infty} |L^k|$.

Let

$$L^k = \begin{bmatrix} 1 & 0 \\ 0 & 1 \end{bmatrix}, \quad \begin{bmatrix} 0 & 1 \\ 1 & 0 \end{bmatrix}, \quad \begin{bmatrix} 0 & 0 \\ 1 & 1 \end{bmatrix}, \quad \begin{bmatrix} 1 & 1 \\ 0 & 0 \end{bmatrix}$$

Then

$$\lim_{k\to\infty} |L^k| = \begin{bmatrix} 1 & 0 \\ 0 & 1 \end{bmatrix}, \quad \begin{bmatrix} 0 & 1 \\ 1 & 0 \end{bmatrix}, \quad \begin{bmatrix} 1 & 0 \\ 1 & 0 \end{bmatrix}, \quad \begin{bmatrix} 0 & 1 \\ 0 & 1 \end{bmatrix}$$

We discard those that do not form a simplex in $\mathbb{R}^1$ with respect to columns, even though the strategy may sample from the strategy simplex at a point. This is to ensure the strategy is undominated, the Markov model holds, and ensure the sample is from the face of the strategy simplex. We are left with:

$$\begin{bmatrix} 1 & 0 \\ 0 & 1 \end{bmatrix}, \quad \begin{bmatrix} 0 & 1 \\ 1 & 0 \end{bmatrix}$$

Recall these matrices of $L_k$ are Nash equilibria. Let $\tau$ be strategies that sample from $\mathcal{M}$ such that:

$$\{\tau \mid \tau \subseteq \prod_{i \in N} h(\omega)\,, \qquad \tau \in \mathbb{R}^2\}$$

Then the following is true:

$$\left\{[\lim_{k\to\infty} \left|L^k\right| \subseteq\ \lim_{k\to\infty} \left|\mathcal{M}^k\right|]\ \cap\ \tau\right\} \subseteq \prod_{i \in N} h(\omega) \rightarrow \left\{\lim_{k\to\infty} \left|\mathcal{M}^k\right| \in \prod_{i \in N} h(\omega) \mid\ (\alpha, \beta) \in \prod_{i \in N} S_i\right\}$$

$$\{\lim_{k\to\infty}|L^k| = \lim_{k\to\infty}|-L^k|\} \cap \tau, \quad L^k \subseteq |\mathcal{M}|, \quad ||L^k||^2 \subseteq ||\mathcal{M}||^2$$

This leads to:

$$\lim_{k\to\infty}|L^k| = L^k, \ \ ||L_k||^2 \subseteq ||\mathcal{M}||^2$$

$$\lim_{k\to\infty}|L_k| \subseteq \lim_{k\to\infty}|\mathcal{M}^k|, \quad \lim_{k\to\infty}|L_k| \cap \tau$$

$$\left[||L_k||^2 \in L_k\right] \cap [\mathcal{M} \cap \tau], \quad ||L_k||^2 \in \tau, \quad ||L_k||^2 \in \prod_{i\in N} h(\omega) \ \blacksquare$$

Theorem 4:

$$\{[\tau \to [\alpha \leftrightarrow \beta]] \downarrow [[\alpha \leftrightarrow \beta] \to \tau]\}$$

Proof: The domain of discourse, initially, is simply Boolean: $\{\tau, \alpha, \beta \mid \tau, \alpha, \beta \in \mathbb{Z}^+, \ [0,1]\}$. Assuming the opposite and converting the disjunction to a conjunction yields the negation of each original disjunct. Converting conditionals to disjunctions, and DeMorgan's law, resolves the rest to get $\sim\tau, \tau$ which is a contradiction and therefore proves the theorem. ■

We then have:

$$\{[\tau \to [[\alpha \to \beta] \cup [\beta \to \alpha]]]\} \bigcap \{[[\alpha \to \beta] \cup [\beta \to \alpha]] \to \tau\}, \text{and} \left\{\alpha, \beta \mid [(\alpha, \beta) \cap \tau] \ \subseteq \prod_{i\in N} h(\omega)\right\}$$

Theorem 5:

$\Gamma_1$ has equilibrium strategies $L_k \subseteq \prod_{i\in N} h(\omega)$ that are a zero-knowledge proof

Proof: By Theorem 2 and Theorem 3, $L_k \subseteq \prod_{i\in N} h(\omega)$. We assume Alice is the verifier and Bob is the prover. Alice challenges Bob regarding one of $\{\alpha, \beta\}$ while also forming an estimate of trust for the opposite component challenged. Furthermore, neither are aware of payoffs. Generalizing from Theorem 2:

$$h_k = \begin{cases} \alpha \otimes \alpha = \alpha \\ \beta \otimes \beta = \beta \\ \alpha \otimes \beta = \begin{bmatrix} 0 \\ 0 \end{bmatrix} \\ \beta \otimes \alpha = \begin{bmatrix} 0 \\ 0 \end{bmatrix} \end{cases}$$

By Theorem 4: $\{\alpha, \beta \mid [(\alpha, \beta) \cap \tau] \subseteq \prod_{i \in N} h(\omega)\}$

We now differentiate between non-revealing strategies and revealing strategies. For informed players, each challenge by Alice should result in a response in the same domain. For Alice to challenge Bob $\alpha$ is contingent on $\beta$ and vice versa. If Bob is uninformed, the best strategy available is to mix strategies according to $\Delta w$ which is the mixed strategy Nash equilibrium under $\prod_{i \in N} h(\omega)$ . At chance, Bob still reveals whether he is informed or not, and indeed mixing under $\Delta w$ with respect to $h_k$ is equal to the choice of chance. Since the choice of chance is equal to $\Delta w$, $L_k$ is non-revealing in that $L_k$ is played when a player is informed with mixture $\Delta w$ and $\widehat{L_k}$ is chosen when Bob is uninformed also with mixture $\Delta w$. However, $h_k(\omega) \in L_k$ always results in $\langle \alpha, \beta \rangle$ and $\widehat{L_k}$ will include the zero vector which demonstrates how informed the opponent is, or over time show the player is uninformed. Additionally, $w\big(h_k(\omega)\big) = \big(\Delta w_\alpha, \Delta w_\beta\big)$. Given non-revealing $L_k$,

$$\lim_{k \to \infty} \rho_I = \lim_{k \to \infty} \big||h_k(\omega)|\big|^2 = w(h_k(\omega)|X) = 1$$

where $\rho_I$ is the probability a player is informed, and under Epsen $\lim_{k \to \infty} \rho_I = \lim_{k \to \infty} 1 - \frac{\frac{2}{3}^k}{1 + \frac{2}{3}^k} = 1$ (Epsen 2006). Then $L_k \subseteq \prod_{i \in N} h(\omega)$ and is a zero-knowledge proof. By Theorem 2, $L_k$ is a Nash equilibrium strategy. ■

Let $\Gamma_2$ be a game $\Gamma_2 = \langle N| \prod_{i \in N} S_i \,|\theta\rangle$ with imperfect recall and incomplete information, with payoff:

$$(X, Y, Z) \in \theta, \;\; X > Y > Z$$

| $\Gamma_2$ | $\alpha$ | $\beta$ |
|---|---|---|
| $\alpha$ | $X$ | $Y$ |
| $\beta$ | $Z$ | $X$ |

*Figure 3 - Game Two*

Let $\Gamma_3$ be a game $\Gamma_3 = \langle N| \prod_{i \in N} S_i \,|\theta\rangle$ with imperfect recall and incomplete information where:

$X, Y, Z \in \theta, X > Y > Z.$

| $\Gamma_3$ | $\alpha$ | $\beta$ |
|---|---|---|
| $\alpha$ | $X$ | $Z$ |
| $\beta$ | $Y$ | $X$ |

*Figure 4 - Game Three*

Theorem 6: $\Gamma_2, \Gamma_3$ are strategically equivalent.

Proof: We adapt the proof of Proposition 4.4 from Boors, et al. (Boors, et al. 2022) to show under permutations of rows and columns in a payoff matrix that each game's payoff matrix can be derived. The relevant definition from Boors, et al. states that payoff bimatrices are strategically equivalent under matrix transformations if a permutation exists with respect to rows, columns, or players (Boors, et al. 2022). The payoff matrices will be defined as:

$$(X, Y, Z \in \theta), \qquad (a, d) \in X, \qquad (b) \in (Y), \qquad (c) \in (Z)$$

$$X > Y > Z$$

The decomposition of the payoff matrices is given by Boors, et al. (Boors, et al. 2022) and adjusted:

$$\theta = \begin{bmatrix} a & b \\ c & d \end{bmatrix}, \qquad \begin{bmatrix} X & Y \\ Z & X \end{bmatrix}$$

$$\begin{bmatrix} X & Y \\ Z & X \end{bmatrix} = \begin{bmatrix} (X) & \left(\frac{Y+Z}{2}, \frac{Y+Z}{2}\right) \\ \left(\frac{Y+Z}{2}\right), \left(\frac{Y+Z}{2}\right) & (X) \end{bmatrix} + \begin{bmatrix} 0 & \left(\frac{Y-Z}{2}, -\frac{Y-Z}{2}\right) \\ \left(-\frac{Y-Z}{2}, \frac{Y-Z}{2}\right) & 0 \end{bmatrix}$$

In each game, permuting the payoffs under these constraints further aligns with Boors, et al. (Boors, et al. 2022), the decomposition between $\theta$ and $\theta'$ are equal:

$$\theta' = \begin{bmatrix} d & c \\ b & a \end{bmatrix}, \qquad \begin{bmatrix} X & Z \\ Y & X \end{bmatrix}$$

$$\text{then for } X > Y > Z \; and \; a > b > c,$$

$$\begin{bmatrix} X & Y \\ Z & X \end{bmatrix} \begin{bmatrix} 0 & 1 \\ 1 & 0 \end{bmatrix} = \begin{bmatrix} Y & X \\ X & Z \end{bmatrix} \rightarrow \begin{bmatrix} X & Z \\ Y & X \end{bmatrix}$$

$$\begin{bmatrix} X & Y \\ Z & X \end{bmatrix} \sim \begin{bmatrix} X & Z \\ Y & X \end{bmatrix}$$

$$\theta \sim \theta'$$

Therefore $\Gamma_2, \Gamma_3$ are strategically equivalent.■

Theorem 7: $\Gamma_2$ has equilibrium strategies $L_k \subseteq \prod_{i \in N} h(\omega)$ that are a zero-knowledge proof

Proof: $w(X) = \frac{1}{2},\ w(Y) = \frac{1}{4}, w(Z) = \frac{1}{4},\ w(h_k|X) = 1$

$$w(h_k) = 2w(X), \qquad w\big(w(h_k(\omega)|X)\big) = w(X) = w(Y) + w(Z) = \frac{1}{2} w(h_k)$$

We adapt Epsen to identify the non-revealing strategy (Epsen 2006):

$$w(h_k) = r_1 w(h_k|X) + r_2 w(h_k|\neg X)$$

$$r_1(h_k) = \frac{r_1(h_k)}{2}, \qquad r_2(h_k) = \frac{r_2(h_k)}{2}$$

$$r_1 = r_2 = w(X) = w(\neg X) = \frac{1}{2} w(h_k) = \frac{1}{2}$$

Then the strategies $L_k$ are non-revealing given they are formed by the choice of chance being equivalent to not knowing the true stage of the game (Epsen 2006). Given non-revealing $L_k$, then $L_k \in \prod_{i \in N} h(\omega)$ and is a zero-knowledge proof. ■

## Fuzzy Zero-Knowledge Model

Linguistic fuzzy-logic game theory does not rely on fuzzy sets, but rather focuses on "computing with words" such that a Nash equilibrium has ordinal preferences (Arfi 2006). Rather than using a truth value, a feasibility value is applied as it has been throughout for $L_k$ (Arfi 2006). Per, Arfi, we have ensured that if this game is reduced from fuzzy-logic, the game is still Boolean and sound per theorems 1 through 7 (Arfi 2006). The feasibility is determined by strings $\{\Phi, \Psi \in \Sigma^*\}$ such that $\Sigma^*$ belongs to a quintuplet adapted from Arfi (Arfi 2006):

$$\left( l \in L_k,\ l^* \in\ L_k^*, \Sigma^*, \Omega, \prod_{i \in N} S_i \right)$$

$l \in L_k$ is then feasibility of the name of the linguistic variable given zero-knowledge signals. In this manner $L_k$ is a domain with a range of linguistic variables that generate the elements of $L_k$, where the domain is composed of zero-knowledge equilibrium strategies as real numbers. In simple terms, the matrix $L_k$ has as elements $\{\alpha, \beta, l\}$ where $\{\alpha, \beta\}$ are 2x1 matrices composed of $l$, where $l$ is the feasibility that $\alpha, \beta$ are true. $L_k$ has been shown to be a non-revealing strategy and mixed strategy Nash equilibrium, and now it includes a range of $\Sigma^*$. $\Sigma^*$ remains unchanged as a language, but we now restrict it to a finite set of words or phrases. Per Arfi, $l^* \in L_k^*$ is defined as a term set. The variable $\omega \in \Omega$ serves as a syntactic rule of being in state $h(\omega)$ which in turn provide the "names of the values" of $l \in$ $L_k$ (Arfi 2006). The mixed strategies serve as the semantic rule for associating the zero-knowledge signaling to the equilibrium strategies, where:

$$\left\{\prod_{i\in N} S_i \cap L_k\right\} \subseteq \Sigma^*$$

The term set $l^* \in L_k^*$ has cardinality relative to partitions such that it is composed of intervals:

$$\{l_0^* = \text{ null}, l_1^* = \text{supported}, l_2^* = \text{confirmed}$$

These values of $l^* \in L_k^*$ and similar are determined by a hedge algebra (Arfi 2006). The hedges are varied modifications which in turn change the meaning of a linguistic structure at a prescribed level (Arfi 2006). It then follows that the hedge algebra results in a DeMorgan lattice with meet, join, negation, and implication operations (Arfi 2006). Here $l^* \in L_k^*$ is the nuance and $l \in L_k$ is the feasibility (Arfi 2006). While the hedges change the primary term, they do preserve a trace of the original meaning (Arfi 2006). This is encapsulated in $\lambda$.

Where $l$ is the feasibility, $l^*$ is the nuance. A nuanced preference is defined as:

$$\left\{\alpha \prec \beta \bigwedge \beta \prec \alpha\right\} \overset{\lambda}{\leftrightarrows} L_k$$

such that $w(X) \prec w(h(\omega)|X)$ for payoff $X$ given likelihood of $h(\omega)|X$ and $X$, and the operator $\wedge$ is the meet operator of the lattice. This defines “crisp orderings of alternative strategies into linguistic orderings” (Arfi 2006):

$$\{w(\neg X) \prec w(X) \prec w(h(\omega)|X)\} \subseteq \Sigma^*$$

This, in addition to linguistic fuzzy-logic conjunction, linguistic fuzzy-logic disjunction, and linguistic fuzzy-logic implication complete the linguistic fuzzy-logic game as a DeMorgan lattice as defined (Arfi 2006). Under this model we see agreement with Arfi in that "rationally choosing a strategy would then consist in optimizing the degrees of nuance and feasibility…" (Arfi 2006).

We now reduce a zero-knowledge game to a fuzzy two-player zero-knowledge game.

Given $\Gamma_4 = \langle N|\Xi\rangle$ for $N$ players and $\xi \in \Xi$ hedges, the following properties are adapted from Arfi to define a linguistic fuzzy-logic game as zero-knowledge (Arfi 2006):

1 – each player has a finite set of choices $\prod_{i \in N} S_i$ where:

$$S_i = \{s_i^o(l, l^*):\ \ l \in \Xi, l^* \in \Xi\}$$

and $l: S_i \to \Xi$ is the feasibility value of choice $s_i^o(l_o, l_o^*)$ given $l_o^*: S_i \to \Xi$ is the nuance degree of $s_i^o(l_o, l_o^*)$. Then $l_i(s_i^o)|o = 1, \dots, q_i$ and $l_i^*(s_i^o)|o = 1, \dots, q_i$ for $q_i$ choices result in a fuzzy logic strategy $\zeta(l_i^o, l_i^{*o})$.

2 - For each player, there is a 2x2 matrix of preference relations which is a subjective ranking of the state space of the game relative to payoff and strategies. Each $L_k, L_k^* \in \mathcal{L}$ degree spans the spectrum of the 2x2 matrix given $\pi|\mathcal{L}| = 1$ for the absolute value of eigenvalues $\pi(\mathcal{L})$.

3 – The game $\Gamma_4$ is defined by 2x2 rules $o \in \mathcal{O}$ such that

$$\mathcal{O}_o = \left\{\left[(s_i^o \uparrow s_{-i}^o) \to \left(o(s_i^o; s_{-i}^o)\right)\right] \mid o_i; o_{-i}; i \in \{1,2\} \neq -i \in \{1,2\}\right\}$$

4 – The strategic arrangement for a player's move $s_i^{o_i}$ is:

$$s_i^{o_i} = o_{o_i o_{-i}}, \qquad \mathcal{O}_{o_i o_{-i}} = \bigcup \mathcal{O}_{k_i k_{-i}}$$

5 – A profile for players in $\Gamma_4$ is defined as the following relations, where players $i, -i$ act relative to rules of inference:

$$GP(i, o_i; -i, o_{-i}) = \bigcap \mathcal{O}_{o_i o_{-i}}$$

Theorem 8: $\Gamma_4$ has an *F-N-Nash Equilibrium* where *F-N-NE* is defined by Arfi (Arfi 2006).

Proof: Given game profile of $\Gamma_4$ with respect to feasibility and nuance, each updated strategy has greater ordinal preference under the hedge algebra such that each move reveals whether a player is informed in increasing accuracy. The rules of inference generating a challenge and response, each attenuated with feasibility and nuance, define how much trust can be placed in an individual. Therefore, each subsequent move has higher ordinal preference until a saturation point. Then the following holds:

$$l_{o_1 o_2}^{pr} \prec l'^{pr}_{o_1' o_2'} \; l_{0_1 o_2}^{*pr} \prec l'^{*pr}_{o_1' o_2'} \blacksquare$$

Theorem 9: There exists an undominated strategy embedded within the strategy simplex.

Proof: $\zeta(l_i^o, l_i^{*o}) \cap L_k$ for fuzzy strategy $\zeta(l_i^o, l_i^{*o})$. Then there exists a *F-N-Nash equilibrium* within $L_k$ under ordinal preferences with respect to the hedge algebra. Determining feasibility of

$$\text{unlikely} \prec \text{less than likely} \prec \text{likely}$$

with nuance:

$$\text{null} \prec \text{supported} \prec \text{confirmed}$$

This allows us to narrow an ordinal interval of acceptable feasibility with appropriate nuance:

$$\text{supported likely} \prec \text{confirmed likely}$$

$$.9 \prec 1$$

Maintaining a left-stochastic matrix:

$$\text{supported likely} \in L_k \to \begin{bmatrix} .9 & .1 \\ .1 & .9 \end{bmatrix}$$

Since $\zeta(l_i^o, l_i^{*o}) \cap L_k$ and $L_k \in \prod_{i \in N} h(\omega)$, $\begin{bmatrix} .9 & .1 \\ .1 & .9 \end{bmatrix}$ is an *F-N-Nash equilibrium* that is non-revealing. ■

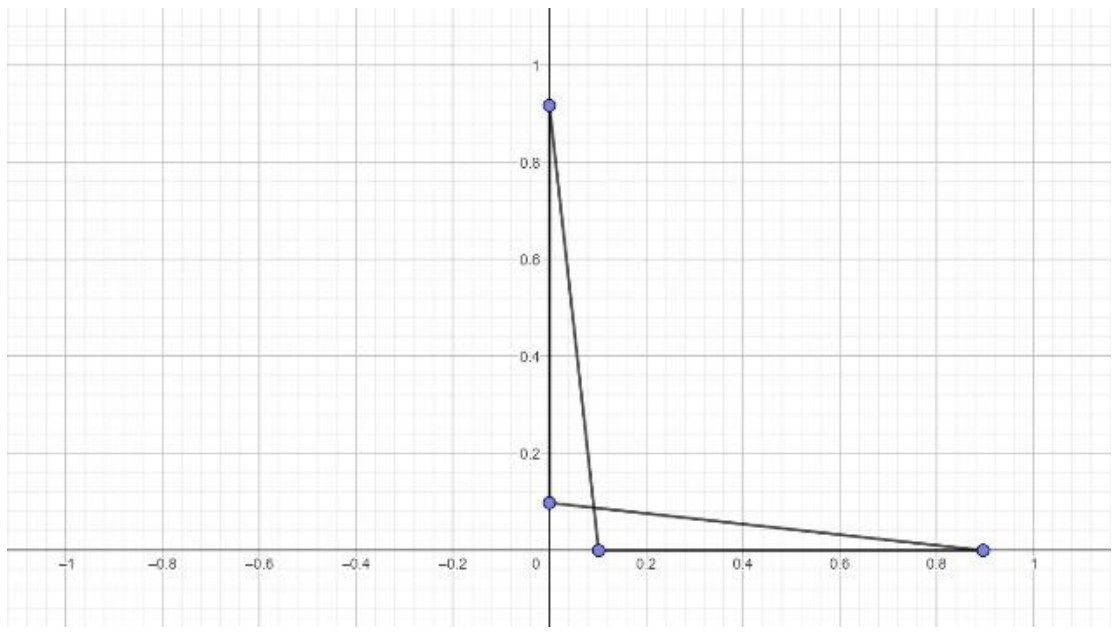

*Figure 5 - Strategy Polytope*

# Conclusion

Comparing Epsen's model of a sample of stage games having zero-knowledge signaling, in terms of this paper (Epsen 2006):

| $G_1$ | | |
|---|---|---|
| | Z | X |
| | Z | Y |

| $G_2$ | | |
|---|---|---|
| | Y | Z |
| | X | Z |

*Figure 6 - Epsen Stage Games*

Under these stage games Epsen showed $w(h_k|G_1) = w(h_k|G_2) = w(h_k) = 1$ for all stage games

$$X > Y > Z$$

(Epsen 2006). We have shown:

$$w(\alpha|\beta) = w(\beta|\alpha) = \lim_{k\to\infty} \left|\left|h_k(\omega)\right|\right|^2 = w(h(\omega)|X) = 1$$

is an equivalent non-revealing strategy for informed players. Furthermore, $\alpha$ can be replaced with $\beta$ and vice-versa, resulting in the equilibrium strategy of choosing what to challenge, and not switching until a player is satisfied their opponent is informed. In this manner we see "commit, confirm, change" and that the strategies themselves are equivalent between zero-knowledge games.

While in $\Gamma_1$ the choice of chance being equal to the mixed strategy NE showed $L_k$ was non-revealing, having a range of mixed strategy NE for $\Gamma_2, \Gamma_3$ with $\frac{1}{2}$ removed is the same as choosing any strategy randomly with $\frac{1}{2}$ removed. Once the payoffs change to a dimension higher than two, the strategies become equivalent and the mixed strategy NE change from a choice of chance and being informed to a choice of any but chance and being informed. The value of $\Gamma_2 \sim \Gamma_3$ symbolized as $\mu$ is given by the equation from Musah et al. (Musah, Kwasi Boah and Seidu 2020):

$$\mu_{2,3} = \frac{ad - bc}{(a + d) - (b + c)} = \frac{X^2 - YZ}{2X - (Y + Z)}$$

For $\Gamma_1$ the value is:

$$\mu_1 = \frac{ad - bc}{(a + d) - (b + c)} = \frac{X^2 - Y^2}{2X - 2Y} = \frac{X}{2} + \frac{Y}{2}$$

The difference in value between the simple two-payoff game vs. higher dimensions of payoff show the precarious costs of zero-knowledge games as defined. For $\Gamma_2$ Alice faces a risk of not knowing whether she incurs a profit or loss while Bob profits. If the game is adjusted under some linear transformation of payoffs, it becomes $\Gamma_3$. Assuming mixed strategy NE of $\Gamma_3$ and understanding each NE is an AE, each party has maximized payoffs. But all payoffs are unknown, and the state is unknown. Since the state is unknown, only a player's strategy is known to them, and the responsive move of their opponent. They will not know much, above chance, but they will know if their opponent is informed.

A thought experiment to understand the game is as follows: Alice and Bob are in the Hilbert hotel. Bob claims he can open every door, and Alice is to verify. Opening an infinite amount of doors would take an infinite amount of time, but if Alice can trust Bob is informed, he need not open every door. A zero-knowledge proof has Alice remain in the lobby while Bob secretly enters rooms and calls the front desk. Alice does not know how Bob enters the rooms, but can challenge the proposition to ask him to enter only even numbered rooms, or every third room. The proof and verification then change accordingly. Over time, Alice will trust that Bob is informed – or not. Alice's payoff is contingent on whether Bob is informed. The more Bob can change the proposition when uninformed to suit his proofs, the higher the cost to Alice. This is essentially called "moving goal posts."

The "will to verify" is critical in a zero-knowledge game. We define the will to verify as a measure of cooperation between all players such that each is acting in good faith with challenges. For Alice being uninformed and challenging Bob, the will to verify plays a large roll given that Alice may challenge either proposition or proof and simply disregard all challenge responses. Eventually Bob will realize Alice is uninformed, but the damage is already done. The more Alice can convince Bob to respond to challenges, Bob may mistake Alice's will to verify as a sign she is acting in good faith. In all cases, damaging or abusing the will to verify costs utility to the malicious actor while costing time for everyone involved. If Bob is uninformed and wishes to degrade the will to verify, he only needs appear informed for small values of $k$. It is true over time under $L_k$ he will appear uninformed, however, if Bob is challenged relative to a proposition, he may alter the proposition enough to justify some value of the truth of the proposition using a proof. In this manner, Bob delays the inevitable but at the same time if he degrades the will to verify, he may appear more informed than he is.

In addition to the "will to verify" is the difference between faithful and faithless actors. In mathematical terms, a subverted game is an inverse function leading to orthogonal projections of which the new coordinate system is the domain of the game.

$$f^{-1}(x,y) = f(x,y)^{\perp} \text{ or } f^{-1}(x,y) = g(x,y,z) \text{ where } g^{-1}(x,y,z) = \begin{cases} f^{-1}(x,y), & \text{for } g = f \\ h(u,x,y,z), & \text{for } g \neq f \end{cases}$$

While malicious actors that are faithless subvert games in their own interest, it is always optimal to subvert a game with arbitrary rules. To determine whether someone is faithful or faithless is different than determining whether a given player is informed. In a zero-knowledge game, determining whether someone is faithless or faithful requires understanding disposition. While understanding disposition is subjective, it can be represented as a range of values as a hedge algebra:

$$\text{Low} \prec \text{Moderate} \prec \text{Favorable}$$

Public announcements are directly proportional to disposition. Every signal is public knowledge, and while players forget moves, under linear transformations the impressions are encoded into public knowledge. We may treat disposition as a risk assessment relative to strategies, and define a stasis as an *ex post* result given a player or coalition achieves an expected payout disproportionate to the opposing player or coalition and the rules are then held fixed. Contrasting this stasis is flourishing, which occurs under Altruistic Equilibrium for all players with coalitions built on being informed.

An unwilling coalition results when no plays are made by any player, assume $f = g$ such that a trivial NE occurs and no public announcements are made. This is a lesser form of flourishing. While it is a stasis, assuming each player is faithful, the maximum payout is reached and equal for all players. A willing coalition converges to an equilibrium exponentially faster than an unwilling coalition. Additionally, trust is additive. A willing coalition in indefinite games with finite subgames in a fuzzy zero-knowledge model will excel in optimized equilibrium exponentially faster than a two-player game without zero-knowledge signals.